\def\year{2019}\relax
\documentclass[letterpaper]{article} 
\usepackage{aaai19}  
\usepackage{times}  
\usepackage{helvet}  
\usepackage{courier}  
\usepackage{url}  
\usepackage{graphicx}  
\usepackage{amsmath}
\usepackage{algorithmic}
\usepackage{textcomp}
\usepackage{xcolor}
\usepackage{subfig}
\usepackage{booktabs}
\usepackage{multirow}
\usepackage{comment}
\usepackage[ruled]{algorithm2e}

\title{Distributionally Robust Semi-Supervised Learning for People-Centric Sensing}

\author{Kaixuan Chen,\textsuperscript{1} Lina Yao,\textsuperscript{1} Dalin Zhang,\textsuperscript{1} Xiaojun Chang,\textsuperscript{2} Guodong Long,\textsuperscript{3} Sen Wang\textsuperscript{4}\\
\textsuperscript{1}School of Computer Science and Engineering, University of New South Wales, Australia\\
\textsuperscript{2}Faculty of Information Technology, Monash University, Australia\\
\textsuperscript{3}Centre for Artificial Intelligence, FEIT, University of Technology Sydney, Australia\\
\textsuperscript{4}School of Information and Communication Technology, Griffith University, Australia\\
}

\begin{document}
\maketitle
\begin{abstract}
Semi-supervised learning is crucial for alleviating labelling burdens in people-centric sensing. However, human-generated data inherently suffer from distribution shift in semi-supervised learning due to the diverse biological conditions and behavior patterns of humans. To address this problem, we propose a generic distributionally robust model for semi-supervised learning on distributionally shifted data. Considering both the discrepancy and the consistency between the labeled data and the unlabeled data, we learn the latent features that reduce person-specific discrepancy and preserve task-specific consistency. We evaluate our model in a variety of people-centric recognition tasks on real-world datasets, including intention recognition, activity recognition, muscular movement recognition and gesture recognition. The experiment results demonstrate that the proposed model outperforms the state-of-the-art methods.
\end{abstract}

\section{Introduction}

People-centric sensing enables a wide range of challenging but promising applications which have great potential on impacting people's daily lives \cite{liao2015human,chen2018interpretable} in many realms such as Brain Computer Interface (BCI) \cite{zhang2018cascade}, assistive living \cite{basanta2017assistive}, robotics \cite{lauretti2017learning} and rehabilitation \cite{smeddinck2015exergames}. One of the major components of people-centric sensing is understanding human behaviors by analyzing the data collected from people-centric sensing devices, such as wearable sensors and biosensors.
However, annotation is difficult in the context of people-centric sensing due to the expensive manual cost, privacy violation and the difficulty in automation \cite{do2014places}. Therefore, a large body of research on semi-supervised learning (SSL) has been proposed. SSL enables a reliable model to be trained by learning from the labeled samples and properly leveraging the unlabeled samples as well. 

Most of the existing SSL works are based on the assumption that the labeled data and the unlabeled data are drawn from identical or similar distributions. 
For example, \cite{cheng2016semi}, \cite{DBLP:conf/ijcai/XingYDWZ18} and \cite{DBLP:conf/ijcai/ChenWGZ18} utilize multiple classifiers to pseudo-label the unlabeled samples that obtain confident predictions. In their tasks, the correctness of labeling is ensured by the condition that the labeled data and the unlabeled data are drawn from similar distributions. But this assumption does not always stand. 

In practical human-centred scenarios, only a few subjects' labeled data can be collected for training and unlabeled data are usually collected from the target users. Since people have diverse behavior patterns and biological phenomena \cite{bulling2014tutorial}, data collected from different subjects are distributed variously. This triggers the distribution shift problem where the labeled data and the unlabeled data are distributed differently. 

Distribution shift is a common problem in people-centric sensing and most practical applications that require predictive modeling. Despite this, the major attention is given to semi-supervised learning of which the main challenge is data scarcity instead of shifted distributions. 
Distribution shift has been relatively underexplored until recently.
Some researchers propose to tackle the distribution shift problem by unsupervised domain adaptation or transferring the model trained on the labeled data to the unlabeled data. For instance, some recent works such as \cite{liu2016coupled}
and \cite{tzeng2017adversarial} are committed to mapping both domains into the common feature space. 
However, they make the covariate assumption that only the marginal distributions of the input data are shifted but overlook the potential shift in the conditional distributions of the output labels given inputs. In this setting, their models only see the difference between the labeled data and the unlabeled data but neglect their latent output-related similarity.

\begin{figure*}[htbp]
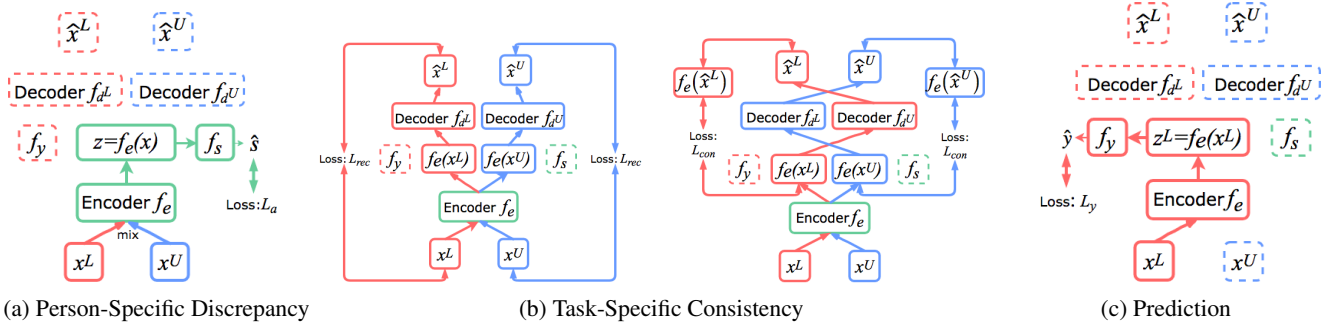

\subfloat[Person-Specific Discrepancy]{
        \includegraphics[width=1.58in]{"Lu"}}
\subfloat[Task-Specific Consistency]{
        \includegraphics[width=1.7in]{"Lrec"}
        \includegraphics[width=1.9in]{"Lcon"}}
\subfloat[Prediction]{
        \includegraphics[width=1.58in]{"Ly"}}
\caption{The overview of the proposed model. We define three components of the training procedure: (a) person-specific discrepancy, (b) task-specific consistency, (c) prediction. Four losses are proposed for the objective: the adversarial loss $L_a$ to reduce person-specific discrepancy, the reconstruction loss $L_{rec}$ and the latent consistency loss $L_{con}$ to preserve task-specific consistency and the prediction loss $L_y$.
When minimizing the losses, only the activated parts are trained (indicated as solid lines) while the rest remain fixed (indicated as dashed lines). Red, blue and green denote the training procedures that are associated with the labeled samples, unlabeled samples, and a mixture of all the training samples, respectively.}
\label{fig:overview}
\end{figure*}

To fill this gap, we propose a two-faced treatment that tackles the problem of SSL for distribution shift.
We define two characteristics for the training data, \textit{person-specific discrepancy} and \textit{task-specific consistency}. Person-specific discrepancy means the distribution divergence of data collected from different people owing to their different behavior patterns and biological phenomena. In our semi-supervised setting, person-specific discrepancy also represents the distribution divergence between the labeled data and the unlabeled data. By contrast, task-specific consistency denotes the inherent similarity of the data of the subjects performing the same task. 
Our aim is to learn an embedding that reduces person-specific discrepancy and simultaneously preserves task-specific consistency.
The main building blocks of the proposed approach are illustrated in Figure~\ref{fig:overview}. We start by reducing person-specific discrepancy. By adversarial training, we reduce the distribution divergence between the latent features of the labeled data and the unlabeled data. Then, we generate paired features and force them to lie in the same space to preserve task-specific consistency. In this way, we ensure the classifier trained with the labeled samples is also effective on the unlabeled samples.

The key contributions of this research are as follows:

\begin{itemize}
    \item We propose a novel distributionally robust semi-supervised learning algorithm to address the distribution shift problem. We consider the distribution discrepancy between the labeled data and the unlabeled data, and align the feature distributions when the training data are distributed differently. We also leverage the similarity of the labeled data and the unlabeled data to learn the task-related discriminative features for classification.
    \item We propose to reduce person-specific discrepancy by aligning the marginal distributions of the labeled data and the unlabeled data. Specifically, we force the latent feature distributions to be similar by training the model in an adversarial way. 
    \item Furthermore, considering the classification task of our model, we propose to preserve task-specific consistency by generating paired data and making their features maintain consistent. Task-specific consistency avoids the features losing the task-related information and facilitates the classification.
    \item 
    We compare the proposed model with eight state-of-the-art methods in four challenging people-centric sensing tasks: intention recognition, activity recognition, muscular movement recognition and gesture recognition. 
    The comprehensive results demonstrate the effectiveness of our model in tackling the distribution shift problem in SSL.
\end{itemize}

\section{The Proposed Method}

\subsection{Problem Statement and Method Overview}

We now detail the distributionally robust model for semi-supervised learning on distributionally shifted data. Assume, there are two parts to the training data: the labeled set $L$ and the unlabeled set $U$. In $L$, each sample ($x^L$, $y$, $s$) consists of an input vector $x^L \in X^L$, an activity label $y \in Y$ and a distribution indicator $s = 1$ that indicates the sample is from $L$, where $X^L$ is some input space and $Y$ is a finite label space for classification problems. In $U$, the samples that lack labels are denoted by ($x^U$, $s$), where $x^U \in X^U$ and $s = 0$ which indicates the sample is from $U$. For simplicity, when referring to a sample regardless of whether it is labeled or unlabeled, we denote the input vector by $x$.

Under the distribution shift assumption, we assume that the data are drawn from different distributions, that is, $L$ is drawn from a marginal distribution $\mathcal{L}(x)$ and $U$ is drawn from a different marginal distribution $\mathcal{U}(x)$. Thus, person-specific discrepancy is formulated as the divergence of $\mathcal{L}(x)$ and $\mathcal{U}(x)$: $Div(\mathcal{L}(x), \mathcal{U}(x))$.
Simultaneously, unlike some domain adaptation methods \cite{liu2016coupled,tzeng2017adversarial} that assume $P_L(y\mid x)=P_U(y\mid x)$, we do not make the same assumption but hold the opinion that there exists latent consistency for data collected in the same tasks. Therefore,
we aim at preserving task-specific consistency by learning latent features $z$ so that $P_L(y\mid z)=P_U(y\mid z)$ and the predictor learned with $L$ is also effective on $U$.

We decompose the proposed model into five parts: an encoder $f_e: X \rightarrow R$ that maps input data to a latent feature $z \in R$, a label predictor $f_y: R \rightarrow Y$ that maps feature $z$ to the label $y$, a distribution predictor $f_s$ that predicts whether the feature $z$ is mapped from $\mathcal{L}(x)$ or $\mathcal{U}(x)$, and two decoders $f_{d^L}: R \rightarrow X$ and $f_{d^U}: R \rightarrow X$ that reconstruct input vectors of $L$ and $U$. The parameters of the five parts are denoted by $\theta_e, \theta_y, \theta_s, \theta_{d^L}, \theta_{d^U}$, respectively. An overview of the proposed model is shown in Figure~\ref{fig:overview}. 

We define four components of the training objective:  the {\em user adversarial} loss, $L_a$, forces a reduction in the distribution divergence of the latent features of $L$ and $U$; the {\em reconstruction} loss, $L_{rec}$, learns two decoders to reconstruct input vectors $\hat{x}$ from latent features $z$; the {\em latent consistency} loss, $L_{con}$, is a constraint that avoids losing the task-specific information during training; the final {\em prediction} loss, $L_y$, encourages the encoder to learn discriminative features and ensures a powerful label predictor is trained.
The total loss can be defined as the sum of the four components:
\begin{equation}
\label{eqn:L_total}
    L_{total} = L_a + L_{rec} + L_{con} + L_y
\end{equation}

\subsection{Reducing Person-Specific Discrepancy}
To reduce person-specific discrepancy, we aim at learning features $z$ and making the distributions $\mathcal{L}_z(z) = \{f_e(x;\theta_e)\mid x\sim\mathcal{L}(x)\}$ and $\mathcal{U}_z(z) = \{f_e(x;\theta_e)\mid x\sim\mathcal{U}(x)\}$ similar. Since calculating and controlling the distribution discrepancy is non-trivial, we force the feature extractor $f_e$ to map $X^L$ and $X^U$ to a unified distribution by learning the features whose distributions cannot be distinguished by the distribution classifier. This is constrained by an adversarial loss $L_a$. (see Figure~\ref{fig:overview}(a)) For the binary classification problem, the loss function is defined as:
\begin{equation}
\label{eqn:L_a}
\begin{split}
    L_a = \frac{1}{N^L}\sum_{n=1}^{N^L}log f_s(f_e(x^L_n)) \\ +\frac{1}{N^U}\sum_{n=1}^{N^U}log(1-f_s(f_e(x^U_n))),
\end{split}
\end{equation}
where $N^L$ is the number of labeled samples and $N^U$ is the number of unlabeled samples. Firstly, we need a sufficiently strong classifier to distinguish users from latent features because successfully deceiving a weak classifier does not mean the features are drawn from similar distributions. This step is done by updating $\theta_s$ while \textit{maximizing} Eq.~\ref{eqn:L_a} and fixing $\theta_e$. Meanwhile, we need $f_e$ to learn the features that are unidentifiable for $f_s$. This is done by updating $\theta_e$ while \textit{minimizing} Eq.~\ref{eqn:L_a} and fixing $\theta_s$. Therefore, the optimization of the adversarial loss can be summarized as:
\begin{equation*}
    \min_{\theta_e}\max_{\theta_s}[L_a(x^L, x^U, \theta_e, \theta_s)]
\end{equation*}
Probabilistically, Eq.~\ref{eqn:L_a} can be rewritten as:
\begin{equation}
\label{eqn:L_a2}
\begin{split}
L_{a} &\approx E_{x\sim\mathcal{L}_x}[log f_s(f_e(x))]  + E_{x\sim\mathcal{U}_x}[log (1-f_s(f_e(x)))]\\
&= E_{z\sim\mathcal{L}_z}[log f_s(z)] + E_{z\sim\mathcal{U}_z}[log(1-f_s(z))]
\end{split}
\end{equation}
The maximum of Eq.~\ref{eqn:L_a2} is related to the Jensen-Shannon divergence between $\mathcal{L}_z(z)$ and $\mathcal{U}_z(z)$ \cite{goodfellow2014generative}:
\begin{equation*}
    \max_{\theta_s}[L_a] = -log(4) + 2 JSD(\mathcal{L}_z(z)\parallel\mathcal{U}_z(z))
\end{equation*}
Thus, 
\begin{equation*}
    \min_{\theta_e}\max_{\theta_s}[L_a] = \min_{\theta_e}[JSD(\mathcal{L}_z(z)\parallel\mathcal{U}_z(z))] -log(4),
\end{equation*}
regardless of the constant, the optimization of the adversarial loss can be formulized as the problem of finding the optimized $\theta_e$ so that the discrepancy between $\mathcal{L}_z(z)$ and $\mathcal{U}_z(z)$ is minimized.

\subsection{Preserving Task-Specific Consistency}

By preserving task-specific consistency, we learn features $z$ so that $P_L(y\mid z)=P_U(y\mid z)$.
Intuitively, if there exists a matching sample $(x^U,x^L)$ that belongs to the same label $y$, we only need to make $f_e(x^L) = f_e(x^U)$. However, in our semi-supervised setting, we do not have paired data to assess the latent task-related differences. Instead, we generate paired data using the decoders shown in Figure~\ref{fig:overview}(b). $f_{d^L}$ and $f_{d^U}$ are able to reconstruct input vectors $\hat{x}$ from the corresponding latent features $z$. They can also be regarded as two generators that generate $\hat{x}$ from $z$. Therefore, we generate $\hat{x}^U$ with $x^L$: $\hat{x}^U = f_{d^U}(f_e(x^L))$, and similarly for the reverse: $\hat{x}^L = f_{d^L}(f_e(x^U))$. In this way, we only need to make $f_e(x^L) = f_e(\hat{x}^U)$ and $f_e(x^U) = f_e(\hat{x}^L)$ to ensure task-specific consistency of the paired data.

Firstly, we need two decoders that can reconstruct input vectors $\hat{x}$ from the corresponding latent features $z$. They are optimized as regular autoencoders (see the left of Figure~\ref{fig:overview}(b)):
\begin{multline}
\label{eqn:L_rec}
    L_{rec}=\frac{1}{N^L}\sum_{n=1}^{N^L}\parallel x^L_n-f_{d^L}(f_e(x^L_n))\parallel^2\\+\frac{1}{N^U}\sum_{n=1}^{N^U}\parallel x^U_n-f_{d^U}(f_e(x^U_n))\parallel^2,
\end{multline}
where $\parallel \bullet \parallel$ denotes the distance between vectors. Note that only two decoders are updated when minimizing $L_{rec}$ since $L_{rec}$ may distract the encoder from learning the features that reduce person-specific discrepancy.
Then, task-specific consistency is ensured by the consistency loss as shown in the right of Figure~\ref{fig:overview} (b):
\begin{multline}
\label{eqn:L_con}
    L_{con}=\frac{1}{N^L}\sum_{n=1}^{N^L}\parallel f_e(x^L_n)-f_e(f_{d^U}(f_e(x^L_n)))\parallel^2\\+\frac{1}{N^U}\sum_{n=1}^{N^U}\parallel f_e(x^U_n)-f_e(f_{d^L}(f_e(x^U_n)))\parallel^2
\end{multline}

We finally conduct the prediction. Good prediction performance not only relies on a powerful predictor but also requires discriminative features.
We harness the annotated data to optimize the parameters of both the feature extractor (the encoder) $f_e$ and the predictor $f_y$ as Figure~\ref{fig:overview}(c) shows. We minimize the empirical loss of the labeled samples by minimizing the cross-entropy between the true label probability distribution and the predicted label probability distribution:

\begin{equation}
\label{eqn: L_y}
    L_y = -\frac{1}{N^L}\sum_{n=1}^{N^L}\sum_{m=1}^{M}y_{n}(m)\log f_y(f_e(x^L_n)),
\end{equation}
where $M$ is the number of label classes, and $y_{n}(m) = 1$ if the $n$-th sample belongs to the $m$-th class and $0$ otherwise. $L_y$ ensures the discriminativeness of the features $z$ learned by the encoder $f_e$ and the good classification ability of the predictor $f_y$ for the annotated data.
Reducing person-specific discrepancy and preserving task-specific consistency ensures that the $f_y$ learned with $L$ only is effective on $U$.

\subsection{Training and Optimization}
The training objective is to minimize Eq.~\ref{eqn:L_total}. Nevertheless, the four losses $L_a$, $L_{rec}$, $L_{con}$ and  $L_y$ have respective goals and different associated parameters to learn. The optimization problem can be summarized and jointly trained as:
\begin{equation}
    \min_{\theta_e,\theta_y}\max_{\theta_s}[L_a+L_{con}+L_y], \min_{\theta_{d^L},\theta_{d^U}}[L_{rec}]
\end{equation}
However, in the experiments, we find that a very strong classifier $f_s$ may minimize the feature distribution discrepancy of $L$ and $U$, but it will also distract the encoder from learning discriminative features for prediction. Therefore, we set a threshold $thre_a$ to seek a balance for the min-max game between person-specific discrepancy and discriminativeness. On the other hand, we require rather strong decoders for reconstruction, a threshold $thre_{rec}$ is thus set to guarantee the reconstruction performance. The detailed procedure is shown in Algorithm~\ref{alg:1}.

\begin{algorithm}[!t]
\caption{Training and Optimization}
\label{alg:1}
\begin{algorithmic}[1]

 \REQUIRE the labeled set $L = \{(x^L,y,s)\}$,\\ the unlabeled set $U = \{(x^U,s)\}$, \\the thresholds $thre_a$, $thre_{rec}$.
 
 \ENSURE  the model parameters $\{\theta_e,\theta_y,\theta_s,\theta_{d^L},\theta_{d^U}\}$.

\STATE $\{\theta_e,\theta_y,\theta_s,\theta_{d^L},\theta_{d^U}\} = RandomInitialize()$

\WHILE{training}

\STATE $L_a\leftarrow Eq.~\ref{eqn:L_a}$
\IF{$L_a < thre_a $}
\STATE $\theta_s\leftarrow \theta_s+\frac{\delta L_a}{\theta_s}$
\ENDIF
\STATE $L_{rec}\leftarrow Eq.~\ref{eqn:L_rec}$
\STATE $\theta_{d^L},\theta_{d^U}\leftarrow \theta_{d^L}-\frac{\delta L_{rec}}{\theta_{d^L}},\theta_{d^U}-\frac{\delta L_{rec}}{\theta_{d^U}}$
\IF{$L_{rec} < thre_{rec} $}
\STATE $L_a,L_{con},L_y\leftarrow Eq.~\ref{eqn:L_a},Eq.~\ref{eqn:L_con},Eq.~\ref{eqn: L_y}$
\STATE $\theta_e\leftarrow \theta_e-\frac{\delta (L_y+L_a+L{con})}{\theta_e}$
\STATE $\theta_y\leftarrow \theta_y-\frac{\delta L_y}{\theta_y}$
\ENDIF

\ENDWHILE

\RETURN $\{\theta_e,\theta_y,\theta_s,\theta_{d^L},\theta_{d^U}\}$

\end{algorithmic}

\end{algorithm}

\section{Experiments}
In this section, we evaluate the performance of our proposed method in four challenging people-centric sensing tasks: intention recognition, activity recognition, muscular movement recognition and gesture recognition.
In particular, we first compare our model with both semi-supervised methods that take no account to distribution shift and other domain adaptation state-of-the-art. The experiment results show that our method outperforms these state-of-the-art methods.
Secondly, we perform a detailed ablation study to examine the contributions of the proposed components to the prediction performance. Then we explore the scalability of our model when $L$ and $U$ are associated with multiple subjects. We further present the visualized distributions of the latent features. Lastly, we analyze the model's sensitivity to the two thresholds.

\subsection{Datasets}
\noindent\textbf{Intention Recognition--EEG Dataset} \cite{goldberger2000physiobank}: 
The EEG dataset contains 108 subjects executing left/right fist open and close intention tasks. The EEG data is collected using BCI2000 instrumentation \cite{schalk2004bci2000} with 64 electrode channels and 160Hz sampling rate. Each subject performs around
45 trials with a roughly balanced ratio of the right and the left fist. We randomly choose 10 subjects for evaluation and select the period from 1 second after the onset to the end of one trial.

\noindent\textbf{Muscular Movement Recognition--EMG Dataset} \footnote{http://archive.ics.uci.edu/ml/datasets/emg+dataset+in+lower+\\limb\#}: 
The UCI EMG Dataset in Lower Limb contains 11 subjects with no abnormalities in the knee executing three different exercises for analysis in the behavior associated with the knee muscle, gait, leg extension from a sitting position, and flexion of the leg up. The data is collected by MWX8 datalog from the Biometrics company. The acquisition process was conducted with four electrodes and one goniometer in the knee. Data with 5 channels are acquired directly from equipment MWX8 at 14 bits of resolution and 1000Hz frequency.

\noindent\textbf{Activity Recognition--MHEALTH} \cite{banos2014mhealthdroid}: 
This dataset is devised to benchmark human activity recognition methods based on multimodal wearable sensor data.
Three inertial measurement units (IMUs) are respectively placed on 10 participants' chest, right wrist, and left ankle to record the acceleration ($ms^{-2}$), angular velocity (deg/s) and the magnetic field (local) data while they are performing 12 activities. The IMU on the chest also collects 2-lead ECG data (mV) to monitor the electrical activity of the heart. All sensing models are recorded at a frequency of 50 Hz.

\noindent\textbf{Gesture Recognition--Opportunity Gesture} \cite{roggen2010collecting}: This dataset consists of data collected from four subjects by a wide variety of body-worn, object-based and ambient sensors in a realistic manner. There are a total of 17 gesture classes that comprises the coarser characterization of the user's hand activities such as opening a door and closing a door, toggle switch. Each recording contains 242 real-value sensory readings.

\subsection{Experiment Setting}
In this work, we use a convolutional autoencoder as the main architecture. The encoder has one convolutional layer, one max-pooling layer and one fully-connected layer. Two decoders use a mirrored architecture with the encoder, including one fully-connected layer, one un-pooling layer and one deconvolutional layer. Each convolutional layer is followed by a rectified linear unit (ReLU) activation and the classification outputs are calculated by the softmax functions. 
The kernel size of the convolutional layer and the deconvolutional layers is $M$ $\times$ $45$ and the number of feature maps is 40, where $M$ denotes the number of features of the datasets and the pooling size is $1$ $\times$ $75$.
We use stochastic gradient descent with Adam update rule to minimize the loss functions at a learning rate of $1e$-$4$. Dropout regularization with a keep probability of $0.5$ is applied before the fully-connected layers. Batch normalization during training is also used to get better performance. All the experiments are conducted on a Nvidia Titan X Pascal GPU.

\subsection{Comparison with State-of-the-Art}

\begin{table*}[!ht]
\centering
\caption{The prediction performance of the proposed approach and other state-of-the-art methods. CNN is a supervised baseline trained with labeled data only. * denotes the domain adaptation state-of-the-art and the others are conventional semi-supervised methods. The best performance is indicated in bold.}
\label{tab:comparison}

\begin{tabular}{c|cccccc}
\hline
\multirow{8}{*}{EEG}         & Method   & CNN         & MSS         & DP          & Tri-Net     & DANN*                \\ \cline{2-7} 
                             & Accuracy  & 66.62$\pm$0.83 & 63.22$\pm$0.64 & 67.90$\pm$1.33 & 67.69$\pm$0.74 & 70.79$\pm$0.81          \\
                             & Precision & 65.46$\pm$0.94 & 58.38$\pm$0.84 & 63.27$\pm$0.84 & 62.36$\pm$0.78 & 69.34$\pm$0.82          \\
                             & Recall    & 67.47$\pm$0.75 & 64.45$\pm$0.77 & 68.61$\pm$0.92 & 68.45$\pm$1.32 & 71.87$\pm$0.88          \\ \cline{2-7} 
                             & Metrics   & CYCADA*     & ADDA*       & CoGAN*      & CycleGAN*   & \textbf{Ours*}       \\ \cline{2-7} 
                             & Accuracy  & 73.29$\pm$0.68 & 67.18$\pm$1.29 & 71.02$\pm$1.27 & 72.69$\pm$0.65 & \textbf{77.01$\pm$0.89} \\
                             & Precision & 73.28$\pm$0.67 & 62.11$\pm$0.77 & 63.83$\pm$0.82 & 63.43$\pm$0.50 & \textbf{73.85$\pm$0.78} \\
                             & Recall    & 73.58$\pm$0.72 & 68.10$\pm$0.84 & 71.94$\pm$0.81 & 73.36$\pm$0.74 & \textbf{75.77$\pm$0.74} \\ \hline
\multirow{8}{*}{MHEALTH}     & Method   & CNN         & MSS     & DP          & Tri-Net         & DANN*                \\ \cline{2-7} 
                             & Accuracy  & 86.67$\pm$0.67 & 87.83$\pm$0.89 & 88.82$\pm$0.89 & 87.07$\pm$0.73 & 89.85$\pm$1.12          \\
                             & Precision & 85.68$\pm$0.85 & 85.53$\pm$0.65 & 86.38$\pm$0.74 & 84.21$\pm$0.68 & 87.56$\pm$1.16          \\
                             & Recall    & 87.06$\pm$0.74 & 85.22$\pm$1.11 & 85.14$\pm$0.6  & 86.30$\pm$0.73 & 90.62$\pm$1.02          \\ \cline{2-7} 
                             & Method   & CYCADA*     & ADDA*       & CoGAN*      & CycleGAN*   & \textbf{Ours*}       \\ \cline{2-7} 
                             & Accuracy  & 92.08$\pm$0.53 & 88.93$\pm$0.68 & 90.35$\pm$0.56 & 91.08$\pm$0.78 & \textbf{95.22$\pm$1.32} \\
                             & Precision & 90.61$\pm$0.53 & 83.72$\pm$1.33 & 86.75$\pm$0.87 & 87.29$\pm$1.32 & \textbf{94.28$\pm$1.16} \\
                             & Recall    & 92.32$\pm$0.97 & 90.35$\pm$0.63 & 90.48$\pm$0.76 & 91.25$\pm$0.86 & \textbf{96.32$\pm$0.86} \\ \hline
\multirow{8}{*}{EMG}         & Method   & CNN         & MSS         & DP          & Tri-Net     & DANN*                \\ \cline{2-7} 
                             & Accuracy  & 64.56$\pm$1.15 & 64.74$\pm$0.68 & 64.78$\pm$0.61 & 66.78$\pm$0.74 & 69.55$\pm$0.68          \\
                             & Precision & 62.11$\pm$0.63 & 63.57$\pm$0.74 & 63.71$\pm$1.08 & 64.23$\pm$0.62 & 66.28$\pm$0.61          \\
                             & Recall    & 66.29$\pm$0.88 & 66.53$\pm$0.70 & 67.61++0.67 & 68.94$\pm$1.32 & 72.15$\pm$0.93          \\ \cline{2-7} 
                             & Metrics   & CYCADA*     & ADDA*       & CoGAN*      & CycleGAN*   & \textbf{Ours*}       \\ \cline{2-7} 
                             & Accuracy  & 74.03$\pm$1.16 & 68.55$\pm$1.11 & 72.37$\pm$0.58 & 74.46$\pm$0.81 & \textbf{77.83$\pm$0.56} \\
                             & Precision & 71.18$\pm$0.92 & 65.83$\pm$0.09 & 70.79$\pm$0.92 & 70.88$\pm$0.61 & \textbf{73.35$\pm$0.74} \\
                             & Recall    & 75.93$\pm$0.53 & 68.10$\pm$0.82 & 73.19$\pm$0.76 & 73.32$\pm$0.67 & \textbf{76.11$\pm$0.65} \\ \hline
\multirow{8}{*}{OPPORTUNITY} & Method   & CNN         & MSS         & DP          & Tri-Net     & DANN*                \\ \cline{2-7} 
                             & Accuracy  & 48.56$\pm$0.62 & 44.15$\pm$0.70 & 47.47$\pm$0.92 & 46.57$\pm$0.84 & 54.82$\pm$0.79          \\
                             & Precision & 49.64$\pm$0.89 & 45.57$\pm$1.32 & 46.85$\pm$0.07 & 45.18$\pm$0.62 & 55.86$\pm$0.73          \\
                             & Recall    & 48.98$\pm$0.74 & 44.16$\pm$0.76 & 44.84$\pm$0.86 & 43.92$\pm$0.96 & 55.90$\pm$0.78          \\ \cline{2-7} 
                             & Method   & CYCADA*     & ADDA*       & CoGAN*      & CycleGAN*   & \textbf{Ours*}       \\ \cline{2-7} 
                             & Accuracy  & 58.83$\pm$0.62 & 52.81$\pm$1.63 & 58.38$\pm$0.75 & 59.23$\pm$1.72 & \textbf{62.27$\pm$0.54} \\
                             & Precision & 58.06$\pm$0.75 & 47.13$\pm$0.98 & 57.30$\pm$1.0  & 53.51$\pm$0.72 & \textbf{60.39$\pm$0.74} \\
                             & Recall    & 58.06$\pm$1.02 & 52.15$\pm$0.91 & 58.31$\pm$0.99 & 60.50$\pm$1.03 & \textbf{62.47$\pm$0.68} \\ \hline
\end{tabular}
\end{table*}

To verify the overall performance of the proposed model, we first compare our model with other state-of-the-art methods. The compared methods include semi-supervised methods (Tri-Net \cite{DBLP:conf/ijcai/ChenWGZ18}, DP \cite{cheng2016semi} and MS \cite{shinozaki2016semi}), none of which take into account distribution shift, and other domain adaptation methods (DANN \cite{ganin2016domain}, CYCADA \cite{hoffman2018cycada}, ADDA \cite{tzeng2017adversarial}, CoGAN \cite{liu2016coupled} and Cycle GAN \cite{zhu2017unpaired}). We also employ a regular CNN as a supervised baseline which is only trained with the labeled set $L$.
Considering that different people have different behavior patterns and biological phenomena, we simulate distribution shift scenarios by drawing training sets $L$ and $U$ from two different subjects $s_L$ and $s_U$. The data of $s_U$ is evenly separated into two, one is the unlabeled training set $U$ and the other is used as the test set $T$.  Cross-validation is conducted on all the participant subjects to ensure rigorousness.

As we can observe from Table~\ref{tab:comparison}, the performance of all the methods on MHEALTH achieves $95\%$ even though $L$ and $U$ are collected from different subjects, while the performance on the other datasets only achieves $60\%$ or $70\%$. The prediction performance demonstrates the degrees of distribution shift in four datasets, among which the discrepancy in MHEALTH is the smallest. This observation coincides with the visualized distribution discrepancy we show in Figure~\ref{fig:visual}.

With respect to the compared methods, the semi-supervised methods Tri-Net, DP and MSS only obtain similar results with regular CNN even though they resort to the unlabeled data of $s_U$. Owing to the distribution shift, the information of $U$ cannot be well leveraged by these methods.
In contrast, DANN, CYCADA, ADDA, CoGAN and Cycle GAN achieve better results since they consider distribution shift and are devoted to mitigating the shift. Overall, the proposed model significantly outperforms the conventional semi-supervised methods. 
Also, our model achieves better performance than other domain adaptation state-of-the-art. By reducing person-specific discrepancy and preserving task-specific consistency, our model makes the classifier $f_y$ trained on $L$ also effective on $U$ and $T$.

\begin{table*}[!ht]
\centering
\caption{Ablation Study. $L_y$ denotes a regular CNN trained with the prediction loss only; $L_y$+$L_a$ is the model trained with a reduction in person-specific discrepancy; $L_y$+$L_{rec}$+$L_{con}$ is the model with preserving task-specific consistency; the last model is our proposed model.}
\label{tab:ablation}
\begin{tabular}{c|ccc|ccc}
\hline
\multirow{2}{*}{Ablation}                                            & \multicolumn{3}{c|}{EEG}                                                                   & \multicolumn{3}{c}{EMG}                                                                  \\ \cline{2-7} 
                                                                     & Accuracy             & Precision            & Recall                                & Accuracy             & Precision            & Recall                                  \\ \hline
$L_y$                                                                & 66.62$\pm$0.83          & 65.46$\pm$0.94          & 67.47$\pm$0.75                    & 64.56$\pm$1.15          & 62.11$\pm$0.63          & 66.29$\pm$0.88                   \\
$L_y$+$L_a$                                                          & 70.79$\pm$0.81          & 69.34$\pm$0.82          & 71.87$\pm$0.88                    & 69.55$\pm$0.68          & 66.28$\pm$0.61          & 72.15$\pm$0.93               \\
\begin{tabular}[c]{@{}c@{}}$L_y$+$L_{rec}$+$L_{con}$\end{tabular} & 68.85$\pm$0.69          & 67.83$\pm$0.86          & 69.35$\pm$0.63                  & 67.40$\pm$0.53          & 65.16$\pm$0.86          & 69.57$\pm$0.56                  \\
\textbf{Our Model}                                                   & \textbf{77.01$\pm$0.89} & \textbf{73.85$\pm$0.78} & \textbf{75.77$\pm$0.74}  & \textbf{77.83$\pm$0.56} & \textbf{73.35$\pm$0.74} & \textbf{76.11$\pm$0.65}  \\ \hline
\multirow{2}{*}{Ablation}                                            & \multicolumn{3}{c|}{MHEALTH}                                                               & \multicolumn{3}{c}{OPPORTUNITY}                                                          \\ \cline{2-7} 
                                                                     & Accuracy             & Precision            & Recall                                 & Accuracy             & Precision            & Recall                                 \\ \hline
$L_y$                                                                & 86.67$\pm$0.67          & 85.68$\pm$0.85          & 87.06$\pm$0.74                  & 48.56$\pm$0.62          & 49.64$\pm$0.89          & 48.98$\pm$0.74                   \\
$L_y$+$L_a$                                                          & 89.85$\pm$1.12          & 87.56$\pm$1.16          & 90.62$\pm$1.02                   & 54.82$\pm$0.79          & 55.86$\pm$0.73          & 55.90$\pm$0.78                  \\
\begin{tabular}[c]{@{}c@{}}$L_y$+$L_{rec}$+$L_{con}$\end{tabular} & 87.41$\pm$1.33          & 86.13$\pm$0.78          & 88.38$\pm$0.91                  & 51.85$\pm$0.92          & 52.23$\pm$0.69          & 54.23$\pm$1.04                    \\
\textbf{Our Model}                                                   & \textbf{95.22$\pm$1.32} & \textbf{94.28$\pm$1.16} & \textbf{96.32$\pm$0.86} & \textbf{62.27$\pm$0.54} & \textbf{60.39$\pm$0.74} & \textbf{62.47$\pm$0.68} \\ \hline
\end{tabular}
\end{table*}

\subsection{Ablation Study}
We perform a detailed ablation study to examine the contributions of the proposed model components to the prediction performance in Table~\ref{tab:ablation}. We first consider the model trained only with $L_y$. This model is composed of $f_e$ and $f_y$, which is the same as a regular CNN trained on $L$ and tested on $T$. This model serves as a baseline to evaluate the effectiveness of the other components. 
Secondly, we evaluate the contribution of reducing person-specific discrepancy by combining $L_y$ and $L_a$. This model is composed of $f_e$, $f_y$ and $f_s$. As we can see in Table~\ref{tab:ablation}, the adversarial loss is effective since the prediction results are improved by $3\%$ to $7\%$. This is in accordance with the analysis that $L_a$ optimizes the parameters of the encoder to minimize person-specific discrepancy and is beneficial to prediction.
We also conduct experiments using the model with preserving task-specific consistency but without the adversarial loss, that is, $L = L_y + L_{rec} + L_{con}$. The model is composed of $f_e$, $f_y$, $f_{d^L}$ and $f_{d^U}$. Note that $L_{rec}$ is only meaningful when it works with $L_{con}$ to build the consistency loop. Otherwise it only trains two decoders of no utilization.
It can be observed that this setting also achieves better performance than the regular model since it directly forces the paired features to be equal and generalizes the model by creating more samples. But it is less effective than reducing person-specific discrepancy. When person-specific discrepancy is large, it is harder to generate data $\hat{x}^L\sim\mathcal{L}(x)$ or $\hat{x}^U\sim\mathcal{U}(x)$ so the effect of preserving task-specific consistency of $x^L$ and $\hat{x}^U$ is limited. When combining all these benefits, our model achieves the best performance.

\subsection{Scalability to Multi-Subjects}

\begin{figure}[!ht]
\centering
\subfloat[EEG]{
        \includegraphics[width=1.6in]{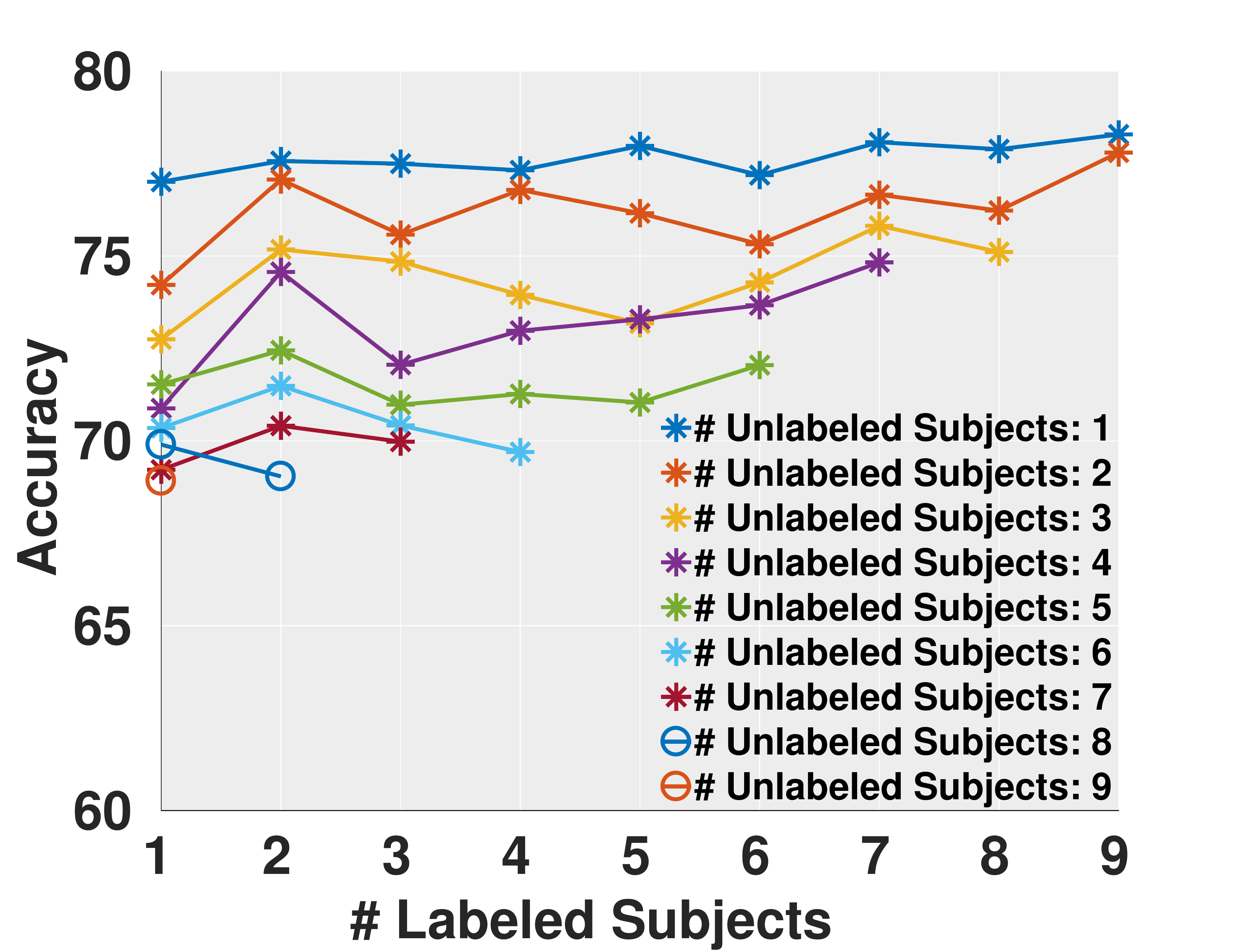}}
\subfloat[MHEALTH]{
        \includegraphics[width=1.6in]{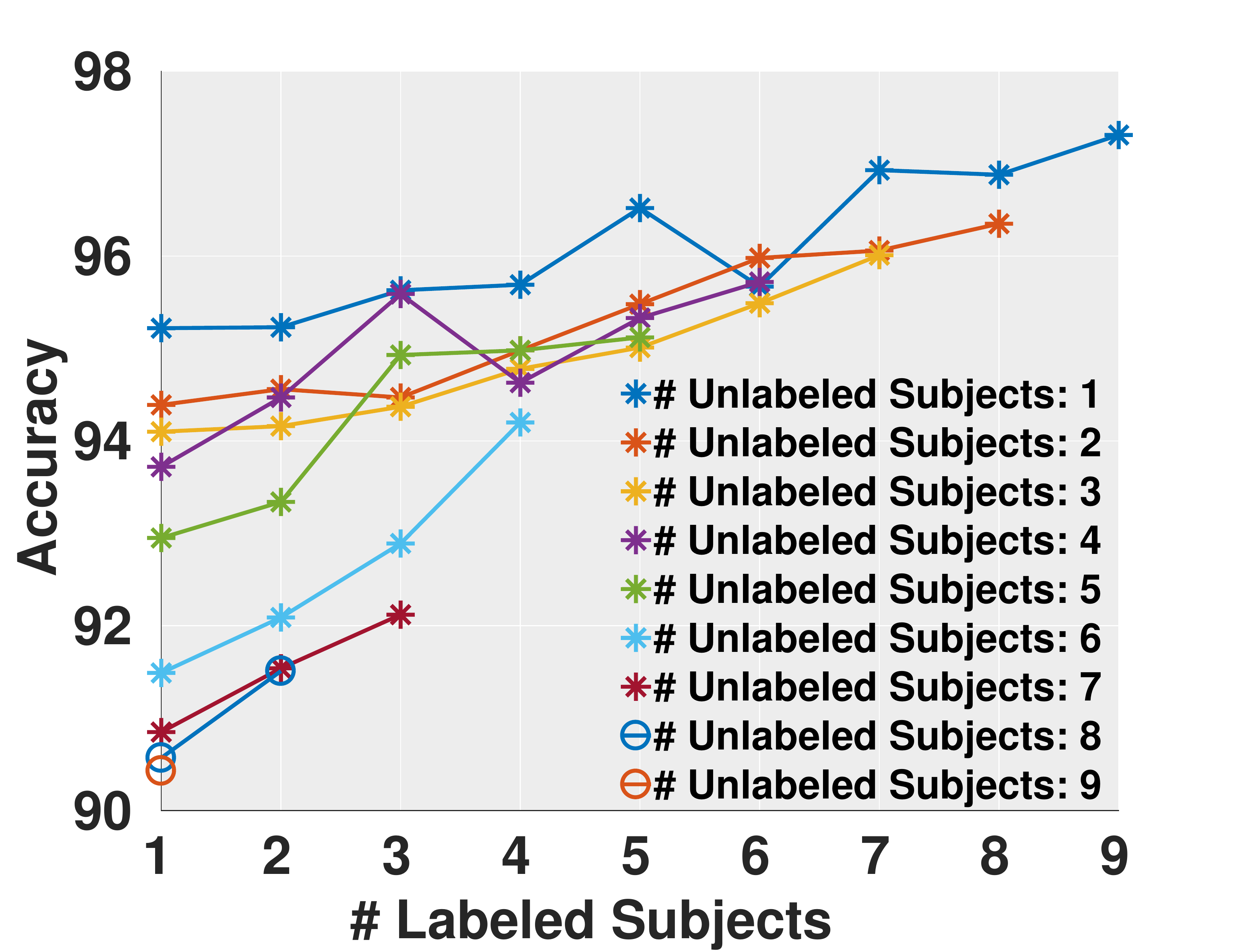}}
        
\subfloat[EMG]{
        \includegraphics[width=1.6in]{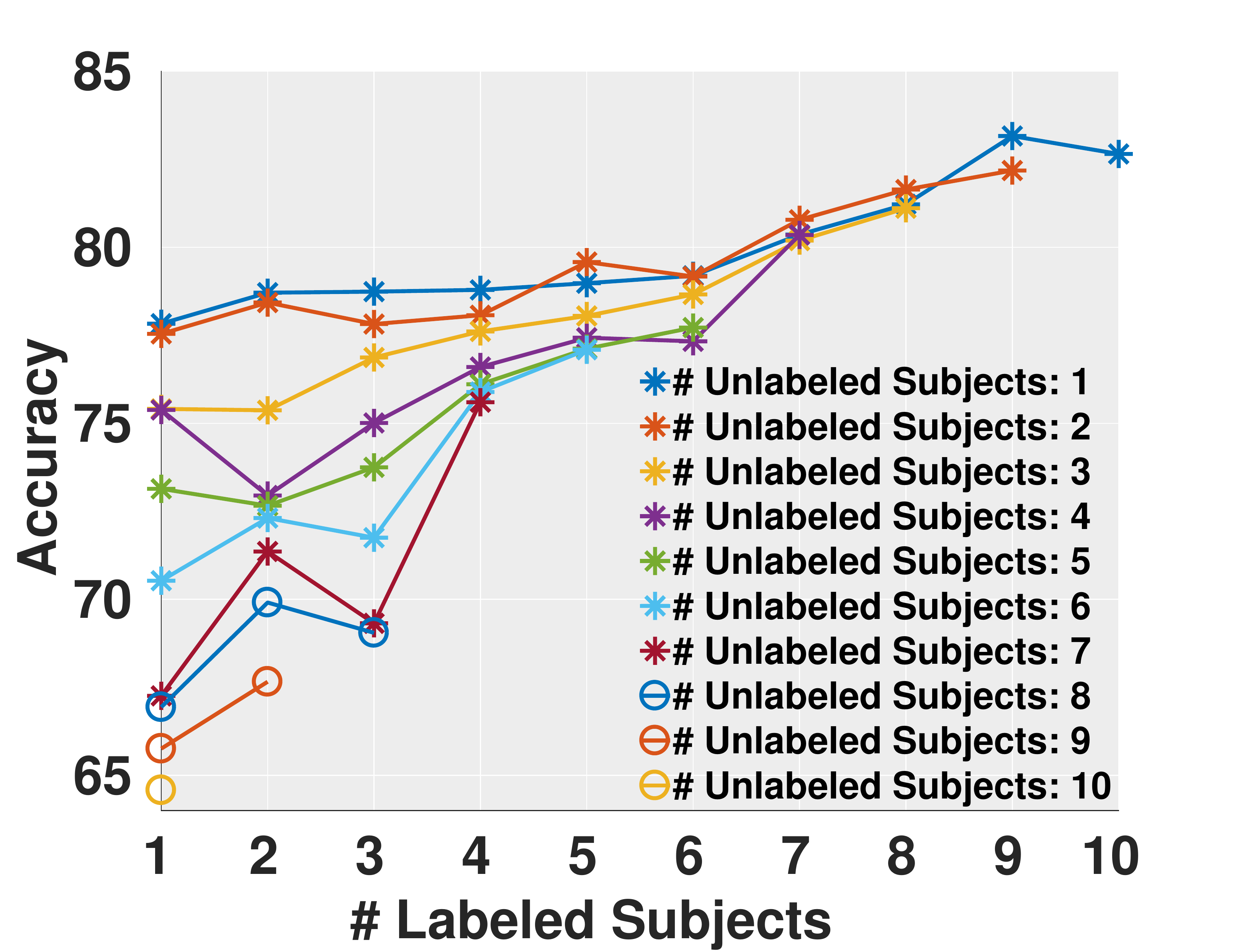}}
\subfloat[OPPORTUNITY]{
        \includegraphics[width=1.6in]{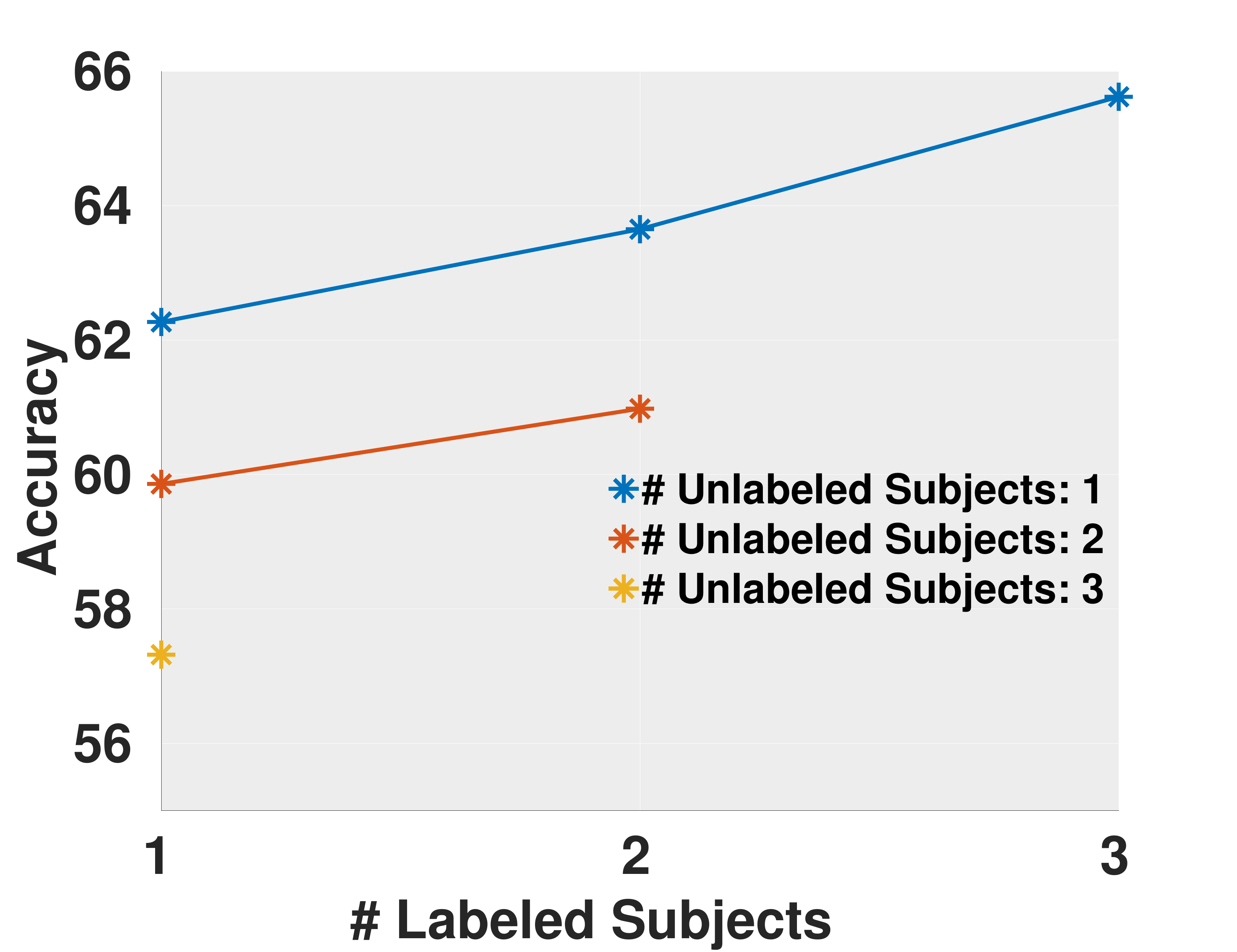}}
\caption{Scalability to Multi-Subjects}
\label{fig:scalability}
\end{figure}

\begin{figure*}[!ht]
\centering
\subfloat[EEG raw]{
        \includegraphics[width=1.7in]{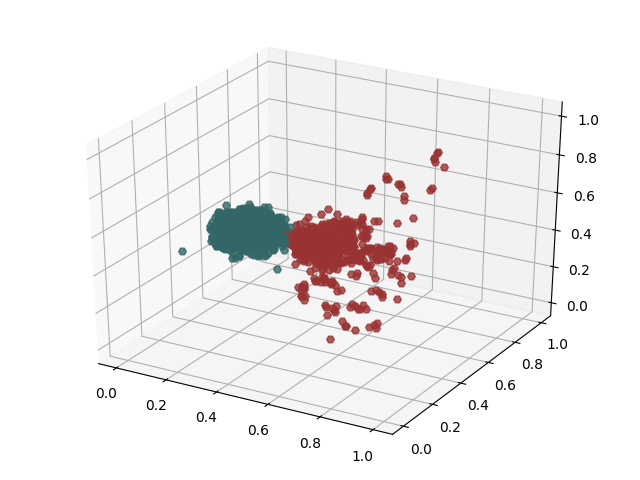}}
\subfloat[MHEALTH raw]{
        \includegraphics[width=1.7in]{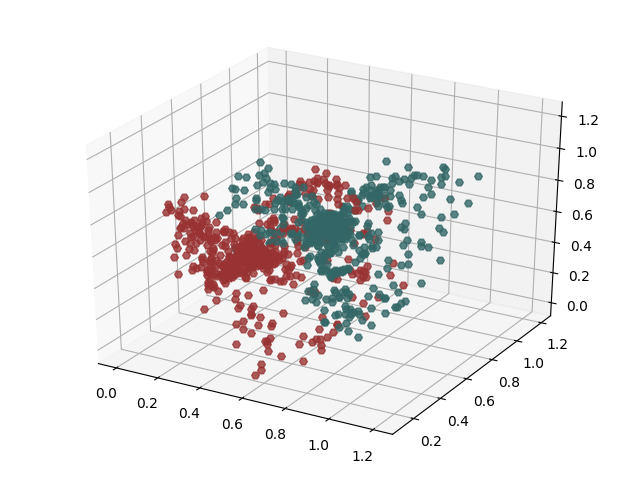}}
\subfloat[EMG raw]{
        \includegraphics[width=1.7in]{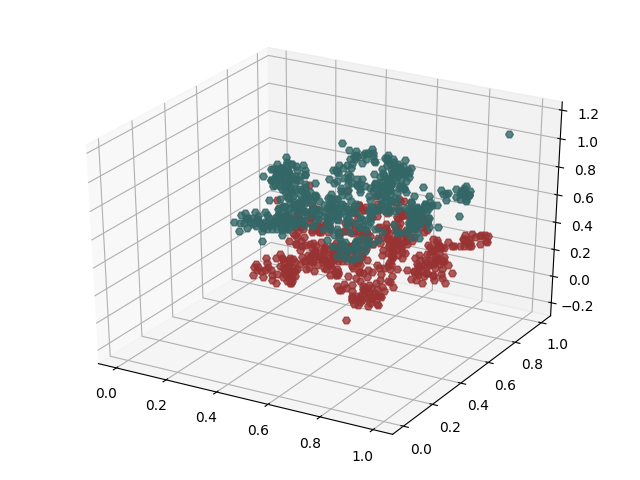}}
\subfloat[OPPORTUNITY raw]{
        \includegraphics[width=1.7in]{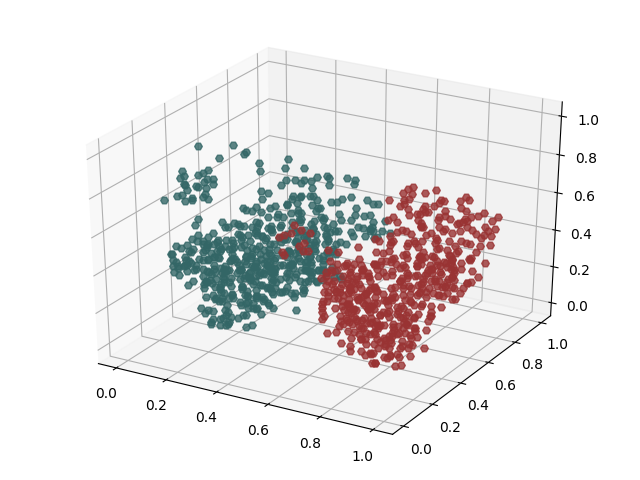}}
        
\subfloat[EEG features]{
        \includegraphics[width=1.7in]{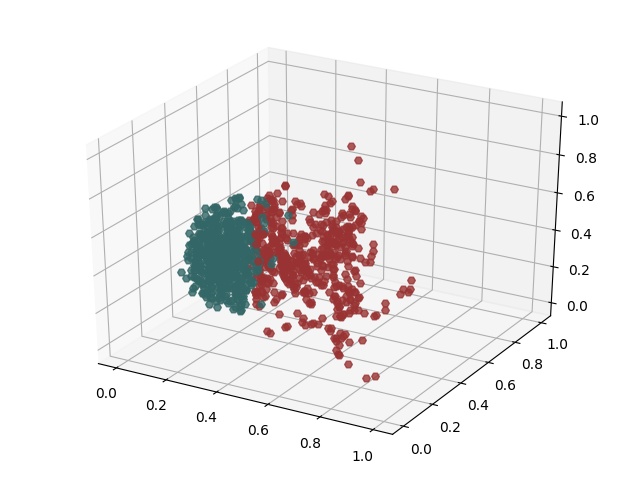}}
\subfloat[MHEALTH features]{
        \includegraphics[width=1.7in]{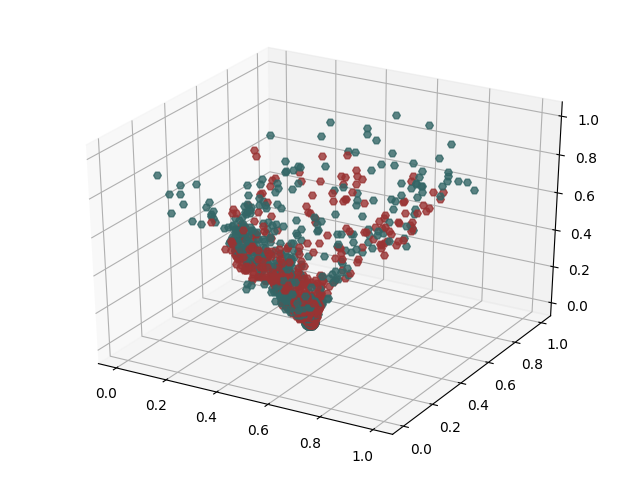}}
\subfloat[EMG features]{
        \includegraphics[width=1.7in]{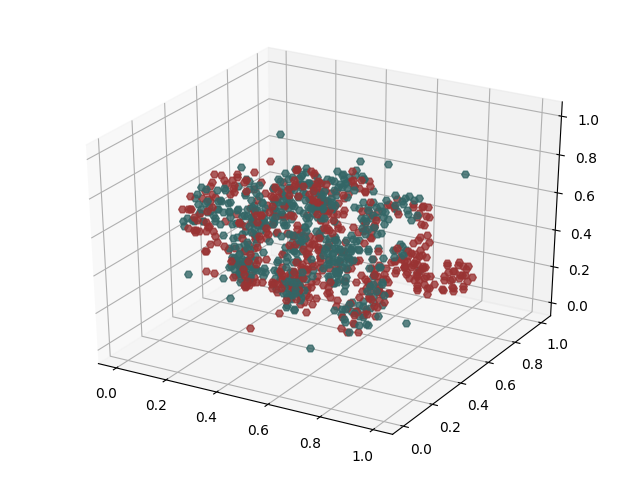}}
\subfloat[OPPORTUNITY features]{
        \includegraphics[width=1.7in]{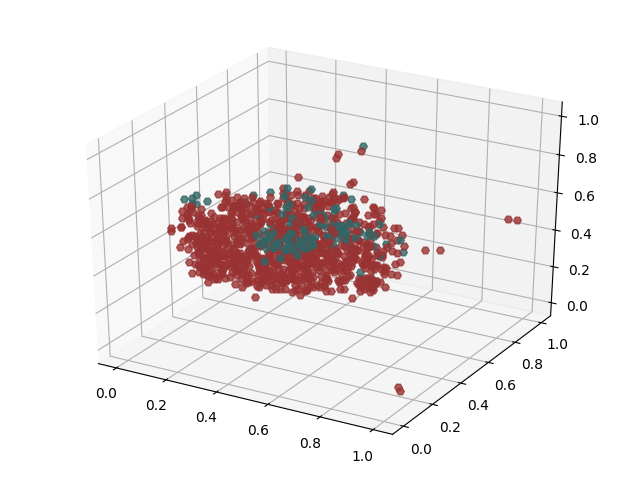}}
        
\caption{Visualization of Latent Features. Green points correspond to the labeled data or features, while the red points correspond to the unlabeled data or features. In all cases, our model is effective in reducing distribution discrepancy.}
\label{fig:visual}
\end{figure*}

The setting of this model is that $L$ and $U$ obey two different distributions. The example is $L$ and $U$ are drawn from two subjects. However, situations still exist when $L$ and $U$ are separately collected from quite a number of subjects. Therefore, the training sets $L$ and $U$ may include multiple diverse distributions. We now explore the scalability of our model in this setting. As Figure~\ref{fig:scalability} shows, we increase the number of the labeled subjects from $1$ to $9$ in the EEG and MHEALTH datasets, from $1$ to $10$ in the EMG dataset, and from $1$ to $3$ in the OPPORTUNITY dataset, and increase the number of the unlabeled subjects in the same way. Note that we do not conduct experiments in the settings when the summation of the number of the labeled subjects and the number of the unlabeled subjects is larger than the total number of the participant subjects since in these settings, there must exist overlapping data shared by $L$ and $U$, which disobeys the overall distribution shift setting. 

In this experiment, the distribution classifier $f_s$ still works as a binary classifier. We consider the merging of all the distributions in $L$ as a new distribution and the same for $U$.
It can be observed that accuracy increases with an increase in the number of labeled subjects and decreases with an increase in the number of unlabeled subjects, which conforms to the intuition that diversely distributed labeled data gives the model generalization ability, but too scattered unlabeled data is detrimental to training.
 
\subsection{Latent Feature Visualization}

To verify the effectiveness of the proposed model, we present the visualized distributions of both the raw data and the latent features of $L$ and $U$ via t-SNE visualization \cite{maaten2013barnes} as Figure~\ref{fig:visual} shows. We can observe a rather obvious discrepancy between the raw data distributions of $L$ and $U$. 
In line with Table~\ref{tab:comparison}, the discrepancy of raw data is relatively unobvious in MHEALTH and is noticeable in OPPORTUNITY.
After training, the features of the labeled data and the unlabeled data are well merged in the MHEALTH, EMG and OPPORTUNITY datasets. The merging is not that effective in the EEG dataset, but a reduction in the discrepancy still can be noticed.

\subsection{Sensitivity to Thresholds}
\begin{figure}[htbp]
\centering
\subfloat[$thre_a$]{
        \includegraphics[width=1.6in]{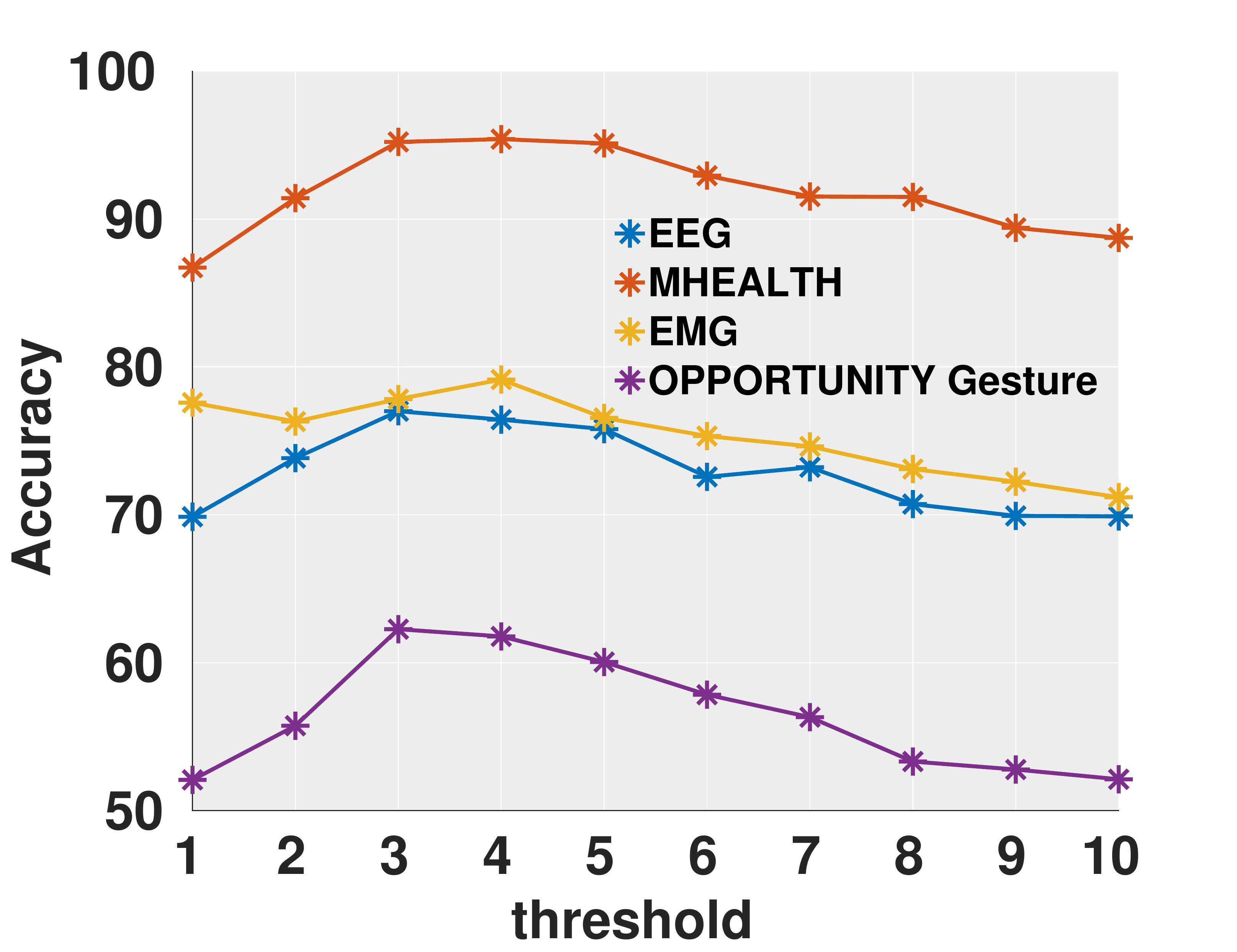}}
\subfloat[$thre_{rec}$]{
        \includegraphics[width=1.6in]{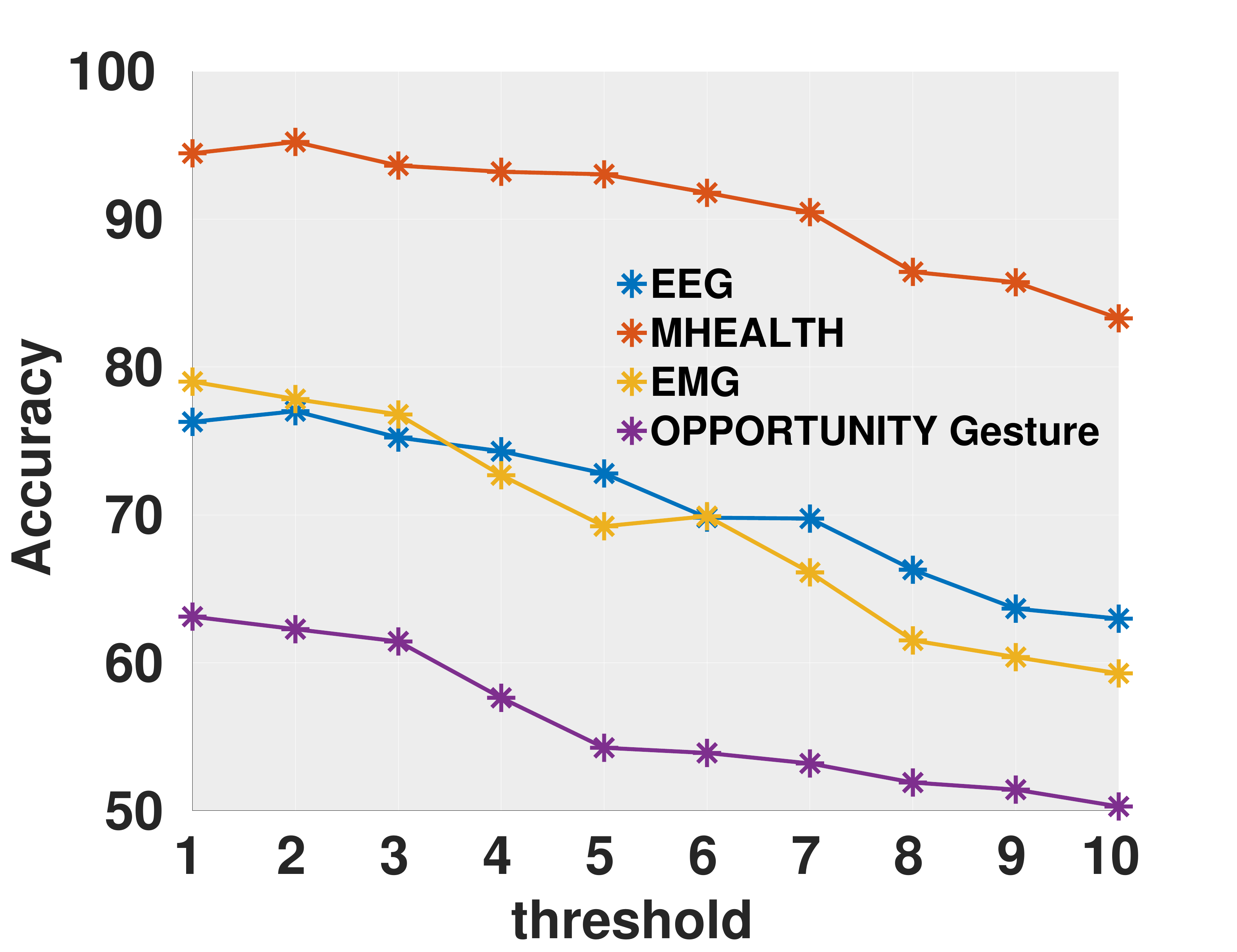}}
\caption{Sensitivity to Thresholds}
\label{fig:thre}
\end{figure}
 Lastly, we present the model's sensitivity to two thresholds in Figure~\ref{fig:thre}.
 $thre_a$ controls how strong the classifier $f_s$ is to align the features of $L$ and $U$, and $thre_{rec}$ affects the reconstruction performance. In Figure~\ref{fig:thre}(a), the prediction accuracy achieves the top when $thre_a$ is around $3$ or $4$. The reason for this is that although a too strong classifier $f_s$ may minimize the feature distribution discrepancy of $L$ and $U$, it also distracts the encoder from learning discriminative features for prediction. Meanwhile, too weak $f_s$ is meaningless to our model. The best $thre_a$, in fact, finds out the balance for the min-max game between person-specific discrepancy and discriminativeness. 
 In Figure~\ref{fig:thre}(b), accuracy decreases with an increase in $thre_{rec}$. It can be inferred that powerful reconstruction ability is significant for the proposed model.

\section{Conclusion}
We propose a novel distributionally-robust semi-supervised method for handling shifted distributions of the labeled and the unlabeled data. The model first reduces person-specific discrepancy by aligning the distributions of the labeled data and unlabeled data. Task-specific consistency is further proposed for extracting label-related features. We experimentally validate our model on a variety of people-centric sensing tasks.
The results demonstrate the outperformance of the proposed model compared with the state-of-the-art. 
Our model is generic and can be applied to practical applications.
\selectfont

\bibliography{chen-yao}
\bibliographystyle{aaai}

\end{document}